\documentstyle[12pt]{article}

  \let\la=\lambda

\let\ph=\varphi  \let\PH=\Phi

\def\0{\over } \def\1{\vec }     \def\2{{1\over2}} \def\4{{1\over4}}
\def\5{\bar }  \def\6{\partial } \def\7#1{{#1}\llap{/}}
\def\8#1{{\textstyle{#1}}}       \def\9#1{{\bf {#1}}}
 \def\llp{\hbox to 0pt{\hss /\hskip1.5pt}}
\def\llo{\hbox to 0.2pt{\hss /}} \def\llq{\hbox to 0pt{\hss /\hskip0.5pt}}
\def\so{\supset\hbox to 0pt{\hss $\displaystyle -$\hskip1pt}}

\def\<{\langle } \def\>{\rangle }

   \let\hc=\dagger

\let\nn=\nonumber
\def\bea{\begin{eqnarray}} \def\eea{\end{eqnarray}}
\def\beann{\begin{eqnarray*}} \def\eeann{\end{eqnarray*}}
\def\beq{\begin{equation}} \def\eeq{\end{equation}}

\date{}
\title{
{\large\rm DESY 95-045}\hfill{\large\rm April 1995}\\
{\large\rm OUTP 95-10 P}\hfill\vspace*{3.0cm}\\
Abelian versus Non-Abelian \\
Higgs Model in Three Dimensions }
\author{W. Buchm\"{u}ller\\
{\normalsize\it Deutsches Elektronen-Synchrotron DESY, 22603 Hamburg,
Germany}\\
\vspace*{0.25cm}\\
O. Philipsen\\
{\normalsize\it Theoretical Physics, University of Oxford,
1 Keble Road, Oxford OX1 3NP, UK}\\
\vspace*{2.0cm}\\
}
\begin{document}

\setlength{\baselineskip}{18pt}
\maketitle
\begin{abstract}
\thispagestyle{empty}
\noindent
We study the phase structure of the abelian Higgs model in three dimensions
based on perturbation theory and a set of gauge independent gap equations
for Higgs boson and vector boson masses. Contrary to the non-abelian
Higgs model, the vector boson mass vanishes in the symmetric phase. In
the Higgs phase the gap equations yield masses consistent
with perturbation theory. The phase transition is first-order
for small values of the scalar self-coupling $\lambda$, where the employed
loop expansion is applicable.
\end{abstract}
\setcounter{page}{0}

\newpage

The ``free-energy functional'' of the Ginzburg-Landau theory of
superconductivity is given by the action of
the abelian Higgs model in three dimensions.
Its phase structure has first been analyzed by Halperin, Lubensky and Ma
\cite{hlm}. For a type-I superconductor, where the scalar self-coupling
$\lambda$ is small compared to the gauge coupling $g$,
the phase transition from the normal ``symmetric'' phase to the
superconducting ``Higgs'' phase is weakly first-order. The case of
a type-II superconductor, where $\lambda/g^2$ is large,
is more complicated
and has been studied by various methods, in particular
the $\epsilon$-expansion and renormalization group techniques \cite{hlm}.

The three-dimensional abelian Higgs model also describes the corresponding
four-dimensional theory at high temperatures. As a model
for the cosmological electroweak phase transition, this case
was studied by Kirzhnits and Linde \cite{kir}, who also found a first-order
transition from the symmetric phase to the Higgs phase for $\lambda/g^2 \ll 1$.
In recent years the abelian Higgs model at high temperatures has been
studied in more detail \cite{arn,bhw} using resummed perturbation theory,
and the effective potential has been determined to order $g^4, \lambda^2$ by a
complete two-loop calculation \cite{heb}.

In the electroweak phase transition non-perturbative effects are expected
to be important, at least for large values of $\lambda/g^2$. They are related
to the infrared behaviour of the non-abelian SU(2) Higgs model in three
dimensions. So far, the nature of the symmetric phase and the order of
the phase transition for large $\lambda/g^2$ have not been firmly established.
In a recent paper \cite{bph} we have studied some non-perturbative aspects
of the SU(2) Higgs model by means of gap equations. Complementing the
mass resummation by a vertex resummation a gauge independent set of
gap equations was obtained for Higgs boson and vector boson masses,
defined on the respective mass shells. The analysis led to the conclusion
that the symmetric phase is again a Higgs phase, just with different
parameters. The first-order phase transition, found for $\lambda/g^2 < 1$,
changes to a crossover at
a critical scalar coupling $\lambda_c$, whose value is correlated with
the magnitude of the vector boson mass in the symmetric phase.

In this letter we apply the same resummation method to the
abelian Higgs model in three dimensions.
Due to the absence of gauge boson self-couplings the abelian
Higgs model does not suffer from the same infrared problems as the
non-abelian theory. It may therefore serve as a testing ground for the
method employed in \cite{bph}.
Much work has been done on the compact and non-compact versions of
the abelian Higgs model on the lattice\footnote{For a review and
references, see \cite{montvay}.}. Monte Carlo simulations
provide evidence for a phase transition from a Higgs phase to a symmetric
Coulomb phase  with zero-mass photon for all values of $\lambda/g^2$
\cite{bar}.

Let us first recall the results of ordinary perturbation theory. Naively,
one may expect that $g^2/m$, $\lambda/m$, $g^2/M$ and $\lambda/M$ appear
as expansion parameters, where $m$ and $M$
denote vector boson and Higgs boson mass,
respectively. In this case the perturbative expansion would fail in the
symmetric phase, for vanishing vector boson mass, $m = 0$. However, due
to the absence of gauge boson self-couplings the infrared behaviour in
the abelian theory is much simpler than in the non-abelian theory.
As shown by Hebecker \cite{heb}, the effective potential does not develop
a term linear in the Higgs field to all orders of perturbation theory in
the case of non-vanishing Higgs mass. The proof immediately implies that the
effective potential is finite to all orders also in the symmetric
phase, with $m=0$ and $M>0$. The convergence of the perturbative expansion
is determined by $g^2/M$ and $\lambda/M$. Following \cite{heb}, one also
easily verifies that $m=0$ to all orders in the symmetric phase.
Hence, perturbation theory in the
symmetric phase is free of infrared divergencies for $M > 0$, and its
results are consistent with non-perturbative studies
on the lattice.

We now use the method developed in \cite{bph} to
derive the gap equations for the
abelian Higgs model. The action of the three-dimensional theory is given by
\beq\label{l3d}
S = \int d^3x \;  \left[{1\over 4}F_{\mu\nu}F_{\mu\nu} +
(D_{\mu}\PH)^\hc D_{\mu}\PH + \mu^2 \PH^\dagger \PH
+  \lambda (\PH^\dagger \PH)^2 \right] \, ,
\eeq
with
\beq
\PH = {1\over \sqrt{2}} (\ph + i \chi) \, ,\quad
D_{\mu}\PH = (\partial_{\mu} - i {g \over 2} A_{\mu})\PH\ .
\eeq
We perform a perturbative calculation in the Higgs phase, i.e.,
we shift the scalar field $\ph$ around its vacuum expectation value
$v$, $\ph=v+\ph'$, and add the $R_\xi$-gauge fixing term
\beq
L_{GF}={1\over 2\xi}(\partial_\mu A_\mu-\xi {g\over 2} v \chi)^2
\eeq
and the corresponding ghost term to the lagrangian (\ref{l3d}).
The shifted lagrangian contains the usual cubic and quartic couplings
between vector field, Higgs field, Goldstone field and ghost field.

At tree level the vector boson, Goldstone boson, ghost and Higgs boson
masses are, respectively,
\beq \label{mtree}
m_0^2={g^2\over 4}v^2,\quad m_{\chi 0}^2=\mu^2+\la v^2+\xi m_0^2, \quad
m_{c 0}^2=\xi m^2_0,\quad M_0^2=\mu^2+3\la v^2 \ .
\eeq
Expanding around the asymmetric tree level minimum one
has $\mu^2+\la v^2=0$, and thus $m_{\chi 0}^2 = m_{c 0}^2 =\xi m_0^2$.
These mass relations aquire corrections in higher orders, and they do
not hold for an expansion around the symmetric minimum $v=0$.

Following the approach of \cite{bph}, we now perform a mass resummation.
The tree level masses are expressed as
\beq\label{masses}
m_0^2 = m^2 - \delta m^2\ ,\ M_0^2 = M^2 - \delta M^2\ ,\
m_{c 0}^2=\xi m^2-\delta m_c^2\ ,\ m_{\chi 0}^2=\xi m^2-\delta m_\chi^2\ ,
\eeq
where the full masses $m$, $M$ and $\sqrt{\xi}m$ enter the propagators
in loop diagrams, and $\delta m^2$, $\delta M^2$, $\delta m_c^2$ and
$\delta m_\chi^2$ are treated as counter terms perturbatively.
Note, that the full ghost and Goldstone boson masses are chosen such
that the tree level mass relations are preserved. Calculation of the
vector boson and Higgs boson self-energies with full propagators
then leads to the coupled set of gap equations
\bea\label{gaps}
\delta m^2 + \Pi_T(p^2 = -m^2, m, M, \xi) = 0\ ,\nn\\
\delta M^2 + \Sigma(p^2 = -M^2, m, M, \xi) = 0\ ,
\eea
where $\Pi_T(p^2)$ is the transverse part of the vacuum polarization
tensor.

As in the non-abelian case, in order to obtain
a gauge independent result for the gap equations (\ref{gaps}),
it is necessary to also perform the following vertex resummations \cite{bph},
\bea\label{vresum}
{g^2\over 2} v &=& g m - \delta V_{\phi\phi\phi}^g \ ,\ \phi = A,\ c,\
 \ph'\ , \chi\ ,\nn\\
\lambda v &=& {g M^2\over 4 m} - \delta V_{\phi\phi\phi}^{\lambda}\ , \
\phi = \ph',\ \chi\ ,\nn\\
\lambda &=& {g^2 M^2\over 8 m^2} - \delta V_{\phi\phi\phi\phi}^\lambda\
,\,
\phi = \ph'\ ,\ \chi\ .
\eea
After these manipulations the lagrangian (\ref{l3d}) takes the form
\beq\label{lresum}
L = L_R + L_1 + L_0\quad .
\eeq
Here, the first term $L_R \equiv L_{RT} + L_{RGF}$ contains
the full masses and vertices which enter
the loop graphs. It is given by the sum of the gauge invariant lagrangian
\cite{bph}
\beq
L_{RT} = {1\over 4}F_{\mu\nu}F_{\mu\nu} + (D_\mu\PH)^{\dagger}D_\mu\PH
    - {1\over 2} M^2 \PH^{\dagger}\PH
    + {g^2 M^2\over 4 m^2}(\PH^{\dagger}\PH)^2\ ,
\eeq
with the Higgs field shifted by the ``classical'' minimum,
$\ph = \ph' + 2m/g$, and the gauge fixing term
\beq
L_{RGF} = {1\over 2\xi}(\partial_\mu A_\mu - \xi m \chi)^2\ ,
\eeq
supplemented by the corresponding ghost lagrangian. $L_1$ in eq. (\ref{lresum})
stands for the difference between tree level and resummed quadratic, cubic
and quartic vertices, and $L_0$ is the constant term of the shifted lagrangian
(\ref{l3d}). $L_1$ and $L_0$ are identical to the expressions given in
\cite{bph}.

For the one-loop self-energies of vector boson and Higgs
boson, as evaluated from the lagrangian $L_R$, we obtain
the result (cf. fig. 1)
\bea \label{vacpol}
\Pi_T(p^2) &=& g^2 \left[{m\over g M^2} v (\mu^2 + \lambda v^2) +
\left({5\over 8}-{M^2\over 8p^2}+{m^2\over 8p^2}\right)
A_0(M^2)\right.\nn\\
&& + \left({1\over 8p^2}(p^2+M^2-m^2)+{m^2\over M^2}\right)
A_0(m^2)\nn\\
&& \left.
+\left({m^2\over 2} -
{1\over 8p^2}(p^2+M^2-m^2)^2\right)B_0(p^2,m^2,M^2)\right]\ ,
\eea
\bea \label{higpol}
\Sigma(p^2) &=& {g^2\over 4}\left[{6\over g m} v (\mu^2 + \lambda v^2)
+3{M^2\over m^2}A_0(M^2)
+\left({M^2\over m^2}+{p^2\over m^2}\right)A_0(\xi m^2)\right.\nn\\
&&
+\left(4-{p^2\over m^2}\right)A_0(m^2)+\left({M^4\over 2m^2}-{p^4\over
2m^2}\right)B_0(p^2,\xi m^2,\xi m^2)\nn\\
&& \left. +{9M^4\over 2 m^2}B_0(p^2,M^2,M^2)
+\left(4m^2+2p^2+{p^4\over 2m^2}\right)B_0(p^2,m^2,m^2)\right]\ ,
\eea
with the three-dimensional integrals
\bea\label{ints}
A_0(m^2) &=& \int {d^3 k\over (2\pi)^3}{1\over k^2 + m^2}\ , \nn\\
B_0(p^2,m_1^2,m_2^2) &=& \int {d^3 k\over (2\pi)^3}
 {1\over (k^2 + m_1^2)((k+p)^2 + m_2^2)}\ .
\eea
The linear divergence of $A_0$ can be cancelled by a counterterm
generated by renormalizing the mass parameter $\mu^2$. The divergence is
absent in dimensional regularization, which we shall use.

Contrary to the non-abelian case, the photon self-energy (\ref{vacpol})
is gauge independent already off the mass shell. However,
the Higgs boson self-energy (\ref{higpol}) has to be evaluated on the mass
shell in order to get a gauge independent result.
Using eqs. (\ref{gaps}) and (\ref{vacpol})-(\ref{ints}), one obtains the
following gap equations for vector boson and Higgs boson masses,
\bea \label{mres}
m^2 &=& m_0^2 - {g z\over M} v (\mu^2 + \lambda v^2)
 + m g^2 \bar{f}(z) \ ,  \\
M^2 &=& M_0^2 - {3g\over 2m} v (\mu^2 + \lambda v^2)
  + M g^2 \bar{F}(z) \ ,\label{Mres}
\eea
with
\bea
\bar{f}(z) &=& {1\over 4\pi}\left[ {1\over 4} +
{1\over 8 z^3} - {1\over 8 z^2} + {1\over 2z} \right. \nn\\
&& \left.\quad +  z^2 - \left({1\over 16 z^4} - {1\over 4 z^2}
+ {1\over 2}\right)\ln(1+2z)\right]\ ,\label{barfw}\\
\bar{F}(z) &=& {1\over 4\pi}\left[({3\over 4}- {9\over 16} \ln3){1\over z^2}
+ {1\over 4z} + z \right.\nn\\
&& \left.\quad - \left({1\over 2}z^2 - {1\over 4} + {1\over 16z^2}\right)
\ln{2z+1\over 2z-1}\right]\ ,\label{barfs}
\eea
and $z=m/M$. For $M>2m$ the equation for $M$ becomes complex, since
in this case the Higgs boson can decay into two vector bosons.

Solutions of the gap equations depend on the vacuum expectation value $v$,
which is defined by the requirement that the expectation
value of the shifted field vanishes,
\beq \label{vevdef}
\< \ph'\> =0\ .
\eeq
This condition on the sum of tadpole graphs yields at one-loop order
in resummed perturbation theory,
\bea \label{vev}
v(\mu^2 + \lambda v^2) &=& - {1\over 4} g m \left( 4 A_0(m^2) +
{M^2\over m^2} A_0(\xi m^2) + 3 {M^2\over m^2} A_0(M^2)\right)\nn\\
&=& {1\over 16 \pi}g\left(4 m^2 + \sqrt{\xi} M^2  + 3 {M^3\over m}\right)\ .
\eea
As in ordinary perturbation theory the vacuum expectation value $v$,
which is not a physical observable, is gauge dependent. This also implies
a weak gauge dependence ${\cal O}(M^2/m^2)$ for
the solutions $m$ and $M$ of the
gap equations. Since the masses are observables, this gauge dependence must
be cancelled by higher order corrections. Numerically, the gauge dependence
of a particular solution of the gap equations
can be used as an indication for the importance of higher order corrections.
In the following we shall work in Landau gauge, $\xi=0$.

For any solution $v$ of eq.~(\ref{vev}) the gap equations (\ref{mres}),
(\ref{Mres}) can be written as
\bea
m^2 &=& {g^2\over 4} v^2 + m g^2 f(z)\ ,\label{mir}\\
M^2 &=& \mu^2 + 3\lambda v^2 + M g^2 F(z)\, ,\label{Mir}
\eea
with functions $f(z)$ and $F(z)$ which can easily be obtained from eqs.
(\ref{barfw}), (\ref{barfs}) and (\ref{vev}). This form of the gap
equations is the same in the abelian and the non-abelian Higgs model,
and it is particularly useful to study the solutions.
In the non-abelian case the function $f(z)$ is positive, except for
very large values of $z$. In particular, for $z={\cal O}(1)$, one has
$f(z) \approx (63 \ln{3} - 12)/(64\pi) \equiv C$ \cite{bph}.
This contribution to $f(z)$ is due to a gauge invariant subset of graphs
corresponding to the gauged non-linear SU(2) $\sigma$-model.
As a consequence, one finds two solutions of the gap equations
for small values of $\mu^2/g^4$ and scalar self-couplings $\lambda$
below a critical coupling $\lambda_c$. One solution,
with $v/g > 1$, corresponds to the usual Higgs phase. The second solution,
with $v/g < 1$, can be interpreted as ``symmetric'' phase, which thus
appears as another Higgs phase with different parameters. To good
approximation the vector boson mass in the symmetric phase is $m = C g^2$.
The range of $\mu^2/g^4$, where two solutions of the gap equations exist,
defines the metastability region where a first-order phase transition
occurs.

In the case of the abelian Higgs model the situation is very different.
Here, the function $f(z)$ is negative for all values of $z$.
This is related to the fact that in the abelian case the non-linear
$\sigma$-model is a free theory. Hence, no
solution with $v/g < 1$ exists, and one is left with a unique solution
of the gap equations
corresponding to the familiar Higgs phase with $v/g > 1$.
This reassures us that the non-trivial values for $v$ and $m$ found
in the symmetric phase of the non-abelian model are not stipulated by our
resummation scheme. It is also consistent
with the fact that in the abelian case
the values $v = 0$, $m = 0$ correspond to
a stationary point of the effective potential to all orders of perturbation
theory, contrary to the non-abelian case! We conclude that in the
abelian Higgs model the trivial vacuum with a massless photon represents
indeed the symmetric phase.

The one-loop results of ordinary perturbation theory can be recovered
from eqs.~(\ref{mres}), (\ref{Mres}) and (\ref{vev}) by substituting
the tree level masses
$m_0=gv/2$ and $M_0=\sqrt{2\la}v$, with fixed ratio
$z=\sqrt{g^2/8\la}$, into the one-loop expressions. This yields
\bea
v(\mu^2 + \lambda v^2) &=& {v^2\over 4\pi}
\left( {1\over 4}g^3 + 3\sqrt{2} \lambda^{3/2}
+ {1\over 2}\sqrt{\xi} \lambda g \right) \ ,\label{pertvev}\\
m^2 &=& -{g^2\over 4\lambda}\mu^2 + {g^3\over 2} v\,
\bar{f}\left(\sqrt{{g^2\over 8 \lambda}}\right)\ ,\\
M^2 &=& -2\mu^2 + g^2\sqrt{2\lambda}v\,
\bar{F}\left(\sqrt{{g^2\over 8 \lambda}}\right)\ .\label{pertM}
\eea
These equations determine the perturbative results for $v$, $m$ and $M$
in the Higgs phase.

{}From the gap equations (\ref{vev})-(\ref{Mir}) the vacuum expectation value
$v/g$ can be obtained as
function of the dimensionless parameters $\la/g^2$ and $\mu^2/g^4$.
In fig.~2 the result is plotted as function of $\mu^2/g^4$,
with $\lambda/g^2 = 1/128$. It agrees well
with the perturbative solution obtained from (\ref{pertvev}).
Also shown is the value $v = 0$, corresponding to the symmetric phase.
For $\mu^2/g^4<0$ the system is in the Higgs phase
with a large vacuum expectation value. This solution of the gap
equations persists up to a small positive value of $\mu^2/g^4$,
where it terminates. The range of small positive $\mu^2/g^4$ with
a Higgs solution \`a la  Coleman-Weinberg \cite{col} corresponds to
the metastability region of the theory. Compared to perturbation
theory, the gap equations predict a smaller range in $\mu^2/g^4$
with metastability. In fig.~3 vector boson and Higgs boson masses
are shown for the same parameters as in fig.~2. In the symmetric phase,
for positive $\mu^2$, the perturbative masses are $m=0$ and
$M=\mu(1 + {\cal O}(g^2,\lambda))$.

Ordinary perturbation theory and also the gap equations are only
reliable for type-I superconductors, where $\lambda/g^2$ and
$M^2/m^2$ are small. As eqs.~(\ref{vev}) and (\ref{pertvev}) show,
the results become strongly gauge dependent otherwise. This indicates
that the one-loop results are no longer trustworthy. For type-II
superconductors other methods have to be used. Particularly interesting
is the use of coarse grained effective actions \cite{wet,pal} where
high frequency modes are integrated out.

For type-I superconductors, with small $\lambda/g^2$,
the gap equations confirm the conventional picture
of a first-order phase transition between a perturbative Higgs phase
and a symmetric Coulomb phase, which is familiar from
ordinary perturbation theory. This result is also in agreement with
non-perturbative numerical simulations on a lattice.
On the contrary, in the non-abelian SU(2) Higgs model a non-vanishing
vector boson mass in the symmetric phase is expected on general
grounds, and it is also found by explicit non-perturbative solutions
of the gap equations. The difference between abelian and non-abelian
Higgs models with respect to the symmetric phase is also reflected
in the nature of the transition.
In the abelian Higgs model one expects a phase
transition for all values of $\lambda$, with a possible change from
first-order to second-order at some critical coupling $\lambda_c$.
For the non-abelian Higgs model, on the other hand, the gap equations
predict a change from a first-order transition to a smooth crossover
already at a rather small value of $\lambda$. Further studies of the
symmetric phase of the non-abelian Higgs model are crucial in order
to achieve a full understanding of the electroweak phase transition.

We are grateful to A. Hebecker, K. Jansen, M. L\"uscher and
M. Teper for valuable discussions and comments.

\section*{Figure captions}
\noindent
{\bf Fig.1a} One-loop contributions to the vector boson propagator.\\
{\bf Fig.1b} One-loop contributions to the Higgs boson propagator.\\
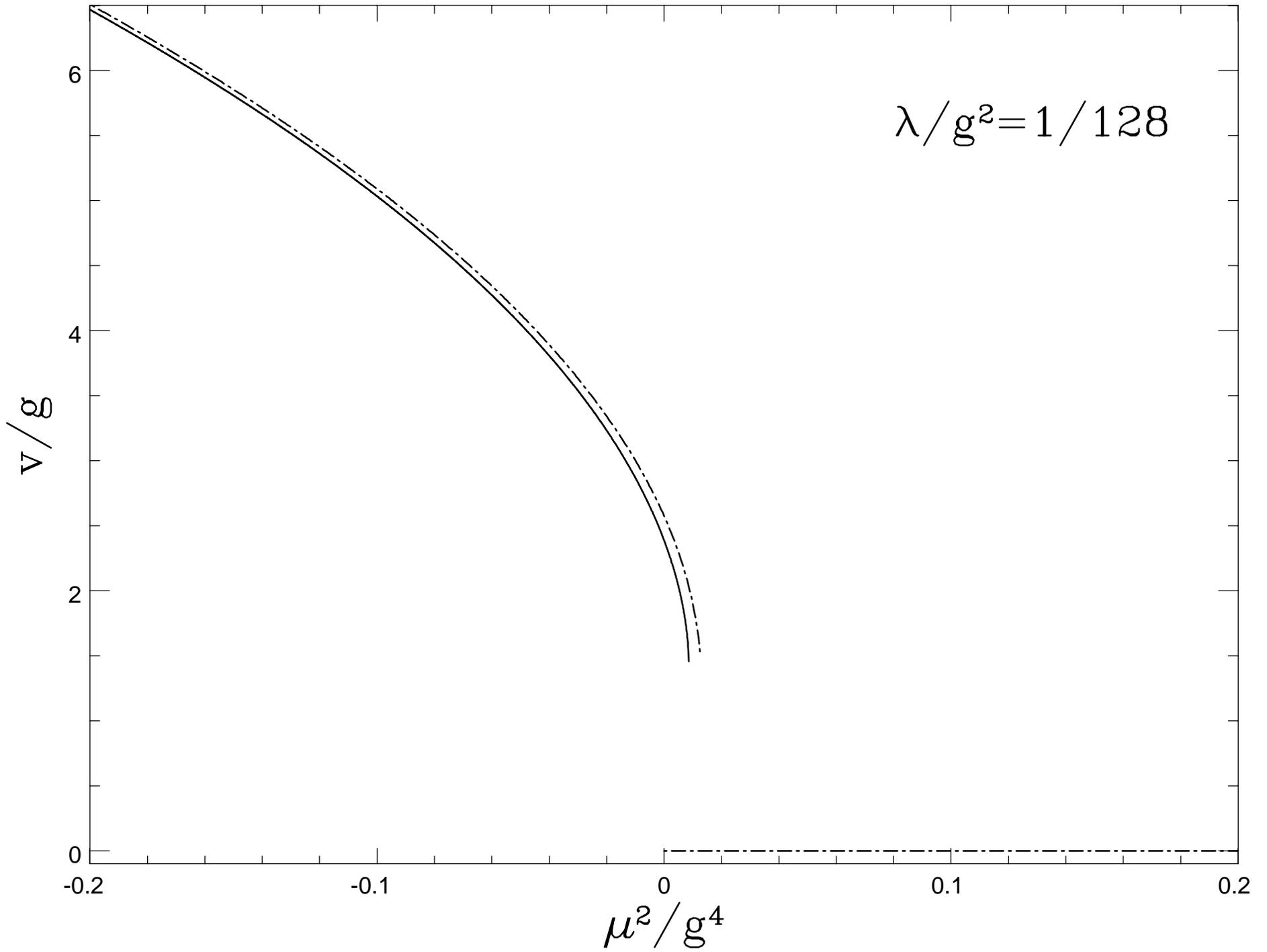
{\bf Fig.2} The vacuum expectation value $v/g$ as function of the
mass parameter $\mu^2/g^4$. Full line: solution of gap equations,
dash-dotted line: perturbation theory. $\lambda/g^2 = 1/128$.\\
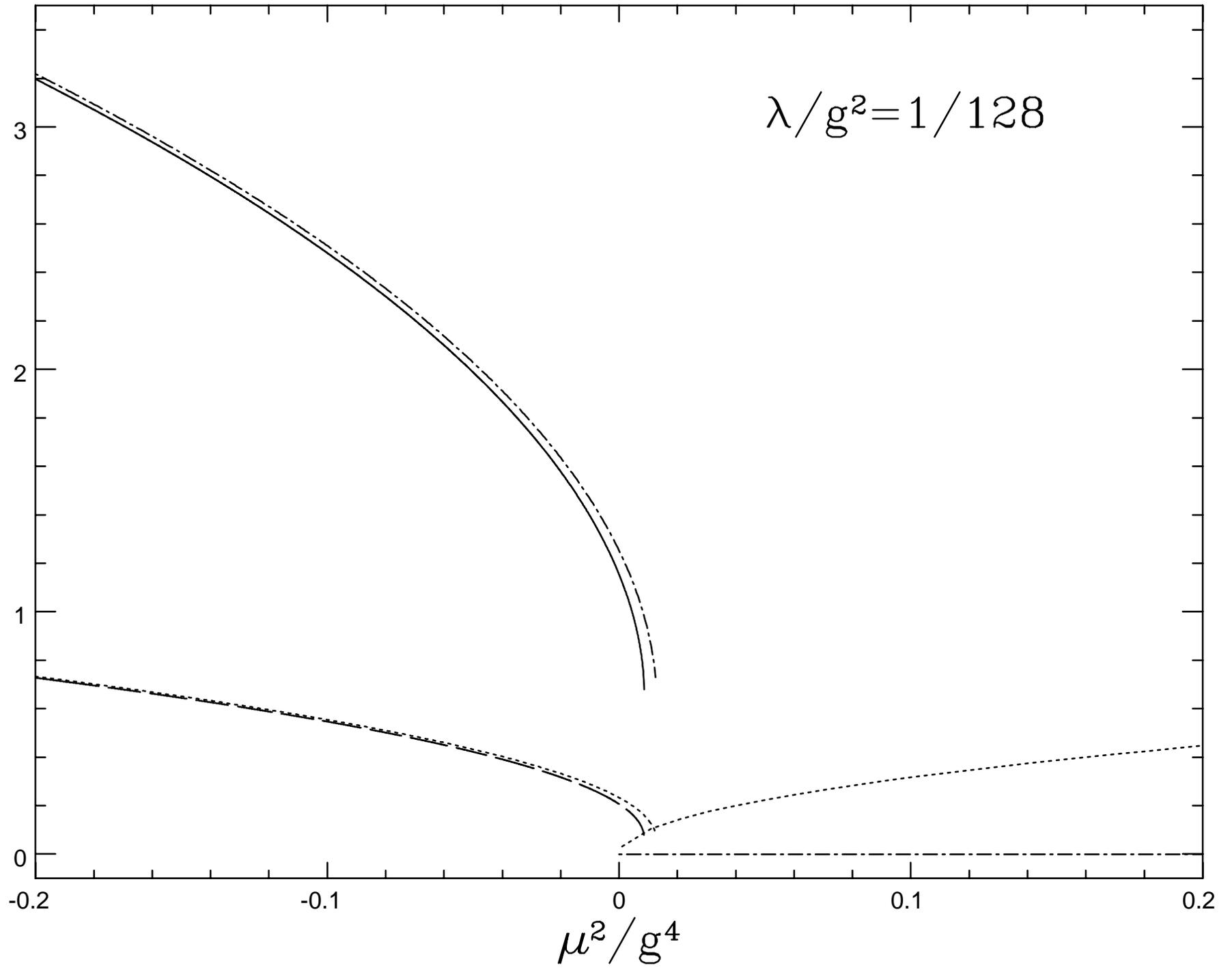
{\bf Fig.3} Vector boson and Higgs boson masses for $\lambda/g^2 =
1/128$.
Gap equations: m (full line), M (dashed line); perturbation theory:
m (dash-dotted line), M (dotted line).
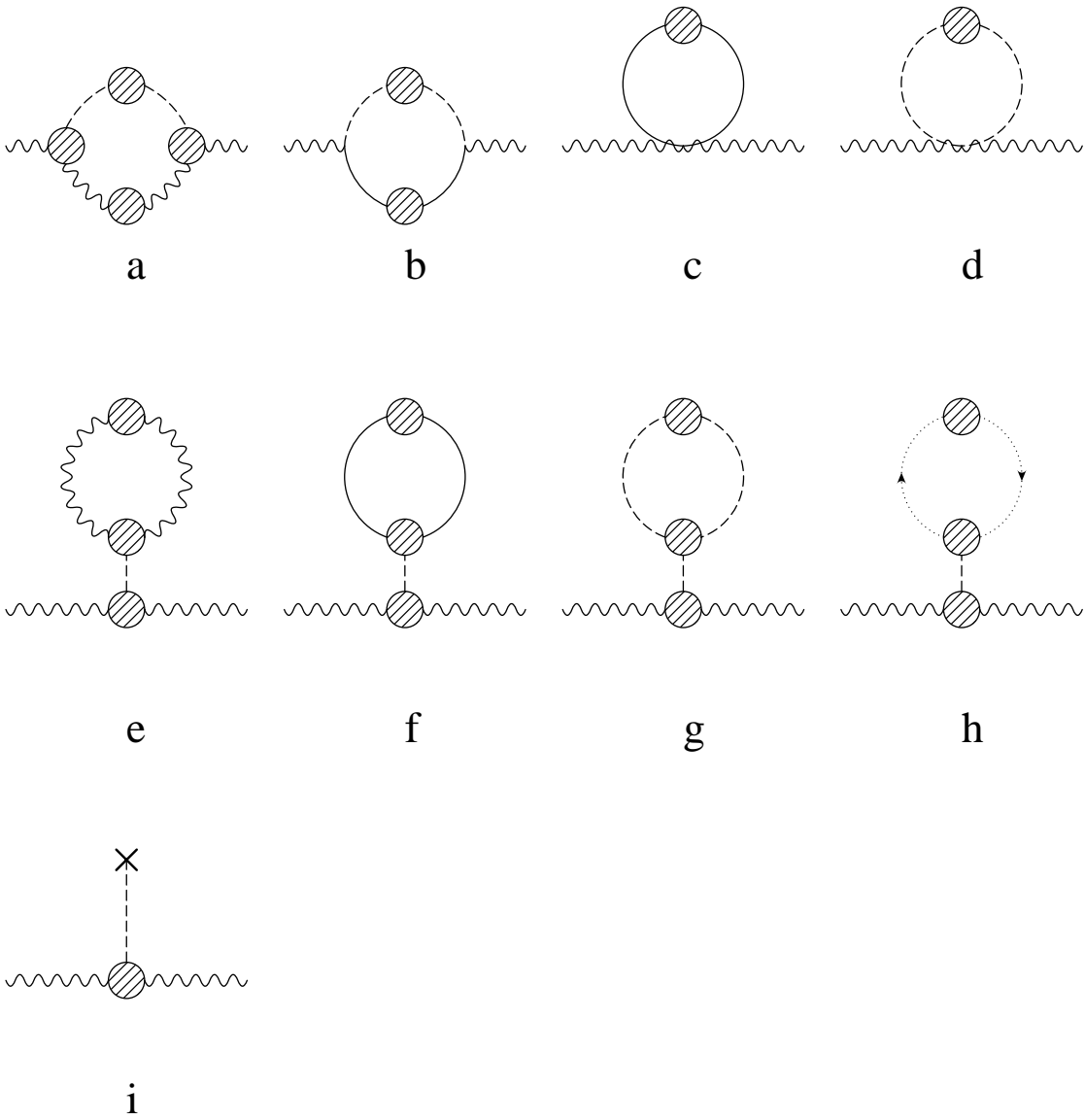
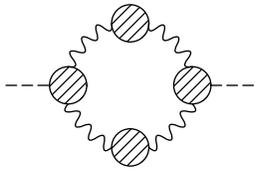
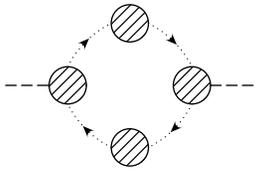
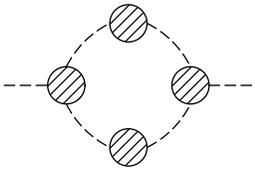
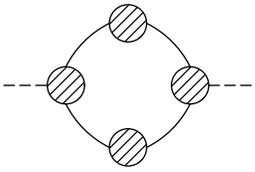
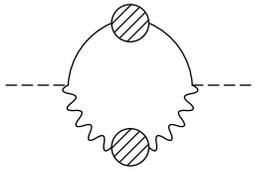
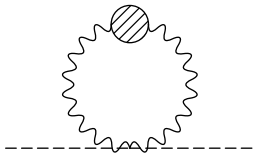
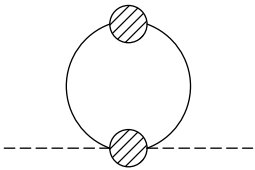
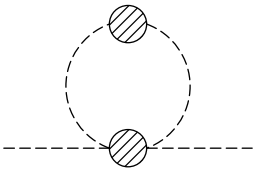
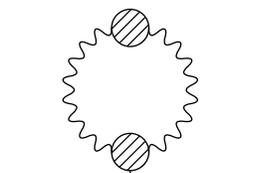
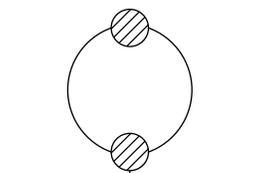
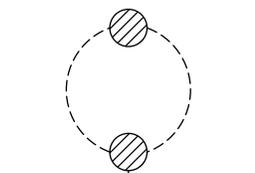
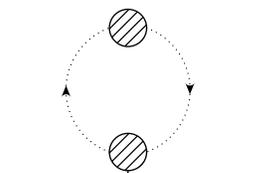
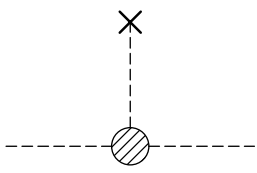
\end{document}